\documentclass{svproc}
\usepackage{amsmath}
\usepackage{amsfonts}
\usepackage{amssymb}
\usepackage{xcolor}
\usepackage{graphicx}
\usepackage{float} 
\usepackage{subfigure} 
\usepackage{cite}

\usepackage{verbatim}
\immediate\write18{texcount -tex -sum  \jobname.tex > \jobname.wordcount.tex}
\usepackage{graphicx}
\title{Review of Recent Advances in  Gaussian Process Regression Methods}
\author{Chenyi Lyu, Xingchi Liu and Lyudmila Mihaylova}
\institute{The University of Sheffield, Sheffield, United Kingdom\\
 Email: \{clyu5, xingchi.liu, l.s.mihaylova\} @sheffield.ac.uk
}
\begin{document}
\maketitle
\begin{abstract}

Gaussian process (GP) methods have been widely studied recently, especially for large-scale systems with big data and even more extreme cases when data is sparse. Key advantages of these methods consist in: 1) the ability to provide inherent ways to assess the impact of uncertainties (especially in the data, and environment) on the solutions, 2) have efficient factorisation based implementations and 3) can be implemented easily in distributed manners and hence provide scalable solutions.
This paper reviews the recently developed key factorised GP methods such as the hierarchical off-diagonal low-rank approximation methods and GP with Kronecker structures. An example illustrates the performance of these methods with respect to accuracy and computational complexity. 
\end{abstract}


\section{Introduction}
Sensors provide huge amounts of data that need autonomous processing and often in real-time. Uncertainty quantification is also necessary for decision-making in autonomous systems. Probabilistic machine learning methods can fill this gap, especially with their data-driven nature and learning capabilities. They can provide a framework for modelling uncertainty that facilitates robust predictions over different changes. Hence, this paper focuses on recent advances in machine learning methods, namely Gaussian Process (GP) regression~\cite{liu2020Gaussian,ki2006Gaussian}. GP is a stochastic process which defines a distribution over possible functions that fit a set of points. This method often produces a fine and precise trade-off between fitting and smoothing the data. 
However, like most machine learning methods, big data can pose challenges to GP methods since it contains much information. Several scalable GP prediction methods have been developed to improve scalability without compromising prediction quality. This paper will review the latest developments in the global GP approximation methods, which approximate the whole probability density function.

Generally, for the global GP approximation, there are two strategies to approximate the kernel matrix with size $n \times n$ through global distillation. The first strategy sets a subset of training data or $m$ $(m \ll n) $ inducing points to construct a low-rank approximation to the GP covariance matrix with a smaller kernel matrix in size $m \times m$~\cite{quinonero2005unifying,quinonero2007approximation}. The capacity of these methods relies on the selection of the inducing point, which requires extra optimisation. Instead of selecting the inducing points, the other strategy uses the low-rank approximation methods, which are based on the spectrum of the covariance matrix to remove the uncorrelated entries in covariance matrix~\cite{shi2011gaussian,ambikasaran2015fast}. Instead of the cube computation cost, most global approximation GP algorithms scale linearly in the training set size. They can offer a close result with the full GP(standard GP without any approximation or factorisation) with enough inducing points~\cite{dai2017efficient}. However, the representation of the global patterns (long-term spatial correlations) often excludes local patterns due to their limited global inducing set~\cite{liu2020Gaussian}. Having appealing properties, GPs have been extensively developed, improved and applied to solve a wide range of artificial intelligence problems.

This paper aims to aid readers in selecting methods appropriate for their applications and further exploring the literature. We present a thorough review of popular global approximation methods, including the methodological characteristics, the advantages, the risks and comparisons for better understanding. Specifically, after a short review of GP regression in Section~2, Section 3 reviews the classic inducing point-based approximation methods. Section~4 contains the Kronecker and Toeplitz-based structured approximation and the structured kernel interpolation (SKI) GP methods. The hierarchical off-diagonal low-rank (HODLR) matrix-based approximation method, which improves GPs in terms of scalability and capability, is reviewed in Section~5. Finally, Section~6 shows the simulation results.

% Section 5~~\cite{ambikasaran2015fast,lyu2022efficient}

\section{Gaussian Process Regression Revisited}
A Gaussian process (GP) is a collection of random variables, any finite number of which have a joint Gaussian distribution. Given any finite set of $n$ inputs $\mathbf{X} =\left\{\mathbf{x}_{1}, \ldots, \mathbf{x}_{n}\right\}$ and the corresponding set of latent function values $\mathbf{f}=\left\{f\left(\mathbf{x}_{1}\right),\ldots f\left(\mathbf{x}_{n}\right)\right\}$. The relationship between the input data $\mathbf{x}_{i}$ and the observed noisy targets $y_{i}$ are given by~\cite{ki2006Gaussian} as
\begin{equation}
    y_{i}=f(\mathbf{x}_{i})+\varepsilon_{i}, \quad \varepsilon \sim \mathcal{N}\left(0, \sigma^{2}\right),
\end{equation}
where $\varepsilon$ is the zero-mean Gaussian noise and $\sigma^{2}$ is the variance of the noise. The prior distribution over the latent function can be written as 
\begin{equation}
    p\left(\mathbf{f} \mid \mathbf{X}\right) \sim \mathcal{N}\left(\overline{\mathbf{f}}, \mathbf{K}\right),
\end{equation}
where the $n \times 1$ mean vector $\overline{\mathbf{f}}$ is defined by mathematical expectation $\mathbb{E}[\mathbf{f}(\mathbf{X})]$. $\mathbf{K}$ is a $n \times n$ covariance matrix which $\mathbf{K}_{i j}=\mathbf{k}\left(\mathbf{x}_{i}, \mathbf{x}_{j}\right)$ are based on the kernel function of GP. The kernel function $\mathbf{k}$ controls the smoothness of the GP.

The predictive distribution~\cite{quinonero2005unifying} of the function values $\mathbf{f}_{*}$ at the test set $\mathbf{X} _{*}$ can be written as
%To predict the test observed noisy target $\mathbf{Y} _{*}$ at $n_{*}$ 
\begin{align}
    p\left(\mathbf{f}_{*} \mid \mathbf{X}_{*}, \mathbf{y}, \mathbf{X}\right) \sim \mathcal{N}\left(\boldsymbol{ \mu}_{*}, \boldsymbol{\Sigma}_{*}\right),
\end{align}
where the mean $\boldsymbol{\mu}_{*}$ and covariance $\boldsymbol{\Sigma}_{*}$ are calculated respectively as
\begin{align}
    \boldsymbol{\mu}_{*}&=\boldsymbol{\bar{\mathbf{f}}}_{\mathbf{X}_{*}}+\mathbf{K}_{\mathbf{X}_{*}, \mathbf{X}} (\mathbf{K}_{\mathbf{X}, \mathbf{X}}+\sigma^{2} \mathbf{I})^{-1}\mathbf{y}, \label{con:fq}\\
    \boldsymbol{\Sigma}_{*}&=\mathbf{K}_{\mathbf{X}_{*}, \mathbf{X}_{*}}-\mathbf{K}_{\mathbf{X}_{*}, \mathbf{X}}(\mathbf{K}_{\mathbf{X}, \mathbf{X}}+\sigma^{2}\mathbf{I})^{-1}\mathbf{K}_{\mathbf{X}, \mathbf{X}_{*}},
    \label{con:cov}
\end{align}
where the $\boldsymbol{\bar{\mathbf{f}}}_{\mathbf{X}_{*}}$ is the $n_{*} \times 1$ mean vector, $n_*$ represents the number of test data, and $\mathbf{I}$ is the identity matrix. $\mathbf{K}_{\mathbf{X}_*,\mathbf{X}}=k(\mathbf{X}_*,\mathbf{X})$ denotes the $n_{*} \times n$ matrix of covariances between the GP evaluated at $\mathbf{X}_{*}$ and $\mathbf{X}$. Also the $\mathbf{K}_{\mathbf{X}, \mathbf{X}}$ is the $n \times n$ covariance matrix evaluated at training inputs $\mathbf{X}$. 
According to~\eqref{con:fq} and \eqref{con:cov}, the computational bottleneck of GP is the inverse of the covariance matrix. For this purpose, a standard procedure is to compute the Cholesky decomposition of $\mathbf{K}_{\mathbf{X}, \mathbf{X}}+\sigma^{2}\mathbf{I}$, requiring $\mathcal{O}\left(n^{3}\right)$ operations and $\mathcal{O}\left(n^{2}\right)$ for storage. Afterwards, the predictive mean and covariance respectively cost $\mathcal{O}(n)$ and $\mathcal{O}\left(n^{2}\right)$ for a single test input.

The marginal likelihood as a function of the model hyperparameters $\boldsymbol{\theta}$ can be utilised to learn the hyperparameters. The negative log marginal likelihood can be written as
\begin{equation}\label{equ:maxlike}
    \log p(\mathbf{y} \mid \boldsymbol{\theta}) \propto-\mathbf{y}^{\top}\left(\mathbf{K}_{\boldsymbol{\theta}}+\sigma^{2} \mathbf{I}\right)^{-1} \mathbf{y}+\log \left|\mathbf{K}_{\boldsymbol{\theta}}+\sigma^{2} \mathbf{I}\right|.
\end{equation}
The first term above evaluates the model fit and the second term penalises the model complexity. By maximising \eqref{equ:maxlike}, the optimal hyperparameters can be learned. However, this maximum likelihood estimation (MLE) for parameter-fitting given noisy observations is a computationally-demanding task since similar to the prediction process \eqref{con:fq} and \eqref{con:cov}, the bottleneck of the adaptation of hyperparameters is the inverse and the determinant of the covariance matrix.

%The prediction $\mathbf{f}=\left[f\left(\mathbf{x}_{1}\right),\ldots f\left(\mathbf{x}_{N}\right)\right]^{\top}$ of a Gaussian process with the new input $\mathbf{X_{*}}$ can be represent as: $p\left(\mathbf{f}\mid\mathbf{X}\right)\sim \mathcal{N}\left(\bar{\mathbf{f}}(\mathbf{x}), k\left(\mathbf{x}, \mathbf{x}^{\prime}\right)\right),$ where the $\bar{\mathbf{f}}$ is the mean function and $k\left(\mathbf{x}, \mathbf{x}^{\prime}\right)$ is the kernel function. Here the mean function is specified as the expectation $\mathbb{E}[.]$:  $\bar{\mathbf{f}}(\mathbf{x})=\mathbb{E}[\mathbf{f}(\mathbf{x})]$ and The kernel function controls the smoothness of GP specified as $k\left(\mathbf{x},\mathbf{x}^{\prime}\right)=\operatorname{cov}\left(f(\mathbf{x})f\left(\mathbf{x}^{\prime}\right)\right)$~\cite{ki2006Gaussian}. Then GP assumes a prior that the latent function behave as:  $\mathbf{x}$ and $\mathbf{x}^\prime$ represent the training and test input data, respectively. The mean function is specified as the expectation $\mathbb{E}[.]$:  $\bar{\mathbf{f}}(\mathbf{x})=\mathbb{E}[\mathbf{f}(\mathbf{x})]$. 

%%%%%%%%%%%%%%%%%%%%%%%%%%%%%%%%%%%%%%%%%%%%%%%%%%%%%%%%%%%%%%%%%%%%%%%%%%%%%%%
\section{Sparse Gaussian Process Approximations}
While the GP approach presented above is powerful, it still faces challenges for large data sets since cubic computing is required. In the last few years, many sparse approximation approaches have been outputted to overcome this limitation~\cite{liu2020Gaussian,quinonero2007approximation}. For most global approximation methods, first define an extra data set $\mathbf{Z} =\left\{z_{1},z_{2}, \ldots, z_{m}\right\}$ in size $m$. The data called inducing points or pseudo inputs, can be mapped to a set of latent inducing variables $\mathbf{f}_{\mathbf{z}}$. These inducing points are introduced to summarise the dependence of the entire training set and achieve sparsity of the full kernel matrix $\mathbf{K}_{\mathbf{X}, \mathbf{X}}$. Based on this, the exact GP can be realised in two different ways: prior approximation and posterior approximation.  

\subsection{Prior Approximation}
Due to the consistency (conjugation) of GPs, the prior $p\left(\mathbf{f}_{*}, \mathbf{f}\right)$ can be recovered by simply marginalizing out $\mathbf{f}_{\mathbf{z}}$ from the joint GP prior $p\left(\mathbf{f}_{*}, \mathbf{f}, \mathbf{f}_{\mathbf{z}}\right)$. Then a fundamental approximation would be introduced which is used in almost all prior sparse approximations. It approximates the joint prior $p\left(\mathbf{f}_{*}, \mathbf{f}\right)$ by presuming that $\mathbf{f}_{*}$ and $\mathbf{f}$ are conditionally independent given $\mathbf{f}_{\mathbf{z}}$, the equation cab be written as
\begin{equation}
p\left(\mathbf{f}_{*}, \mathbf{f}\right) \simeq q\left(\mathbf{f}_{*}, \mathbf{f}\right)=\int q\left(\mathbf{f}_{*} \mid \mathbf{f}_{\mathbf{z}}\right) q(\mathbf{f} \mid \mathbf{f}_{\mathbf{z}}) p(\mathbf{f}_{\mathbf{z}}) \mathrm{d} \mathbf{z},
\end{equation}
where $ p(\mathbf{f}_{\mathbf{z}})=\mathcal{N}\left(\mathbf{0}, \mathbf{K}_{\mathbf{z}, \mathbf{z}}\right)$ and $\mathbf{K}_{\mathbf{z}, \mathbf{z}}$ are the covariances between the including points $\mathbf{Z}$ itself. The prediction function $\mathbf{f}_{*}$ can only connect to $\mathbf{f}$ through inducing points $\mathbf{Z}$. Consequently, there are dependencies between the training and test cases.
In different approaches, there are different additional assumptions between two approximate inducing conditionals $q(\mathbf{f} \mid \mathbf{f}_{\mathbf{z}})$, $q\left(\mathbf{f}_{*} \mid \mathbf{f}_{\mathbf{z}}\right)$. 

Snelson~\cite{snelson2006sparse} proposes a new likelihood approximation method to speed up GP regression called the sparse pseudo input GP (SPGP) method. This model has a sophisticated likelihood approximation. Based on this SPGP method, the training conditional can be taken as fully independent data, which leads to the fully independent training conditional (FITC) approximation method~\cite{saatcci2010Gaussian}. It removes all the dependencies between latent variable function values, which means the values of the latent variable function are conditionally fully independent of given the inducing points $\mathbf{Z}$. By using a diagonal matrix, the FITC prior based on the inducing points can be written as
\begin{equation}
q_{\text {FITC}}(\mathbf{f} \mid \mathbf{f}_{\mathbf{z}})=\mathcal{N}\left(\mathbf{K}_{\mathbf{X}, \mathbf{Z}} \mathbf{K}_{\mathbf{Z}, \mathbf{Z}}^{-1} \mathbf{Z}, \operatorname{diag}\left[\mathbf{K}_{\mathbf{X}, \mathbf{X}}-\mathbf{Q}_{\mathbf{X}, \mathbf{X}}\right]\right),
\end{equation}
where $\operatorname{diag}[.]$ denotes a diagonal matrix and $\mathbf{Q}_{\mathbf{X}, \mathbf{X}} \triangleq \mathbf{K}_{\mathbf{X}, \mathbf{Z}} \mathbf{K}_{\mathbf{Z}, \mathbf{Z}}^{-1} \mathbf{K}_{\mathbf{Z}, \mathbf{X}}$. The prior approximations recover the full GP when the size of inducing points $m=n$. However, this configuration is not the global optimum when maximising $\log q_{\text {FITC}}$, which makes them philosophically troubling. Since the inducing point might be greedily selected one by one, learning inducing points via optimising (6) may produce poor predictions. These issues will be addressed by the posterior approximations reviewed below. However, compared to the posterior approximation, FITC will return better error-bar estimates. This method offers good accuracy and low computational cost. 
%
% I do not understand what you mean by "control group" ???
%which is assumed as a control group in the recent GP papers. 

\subsection{Posterior Approximations}
A sparse GP that utilises inducing variables must select the inducing points and the kernel parameters~\cite{blei2017variational}. Unfortunately, this method has two disadvantages when searching for hyperparameters and inducing points. Due to the change in the prior, the continuous optimisation of the maximum likelihood equation (6) for the location of inducing point $\mathbf{z}$ cannot approximate the GP model. Furthermore, overfitting is possible because the maximum likelihood is highly parametric in this situation (due to the extra hyperparameters as the locations of inducing points). When jointly optimising the matrix $\mathbf{Q}_{\mathbf{X}, \mathbf{X}}$, since the prior has changed, continuous optimisation of maximum likelihood for inducing points $\mathbf{z}$ cannot yield a reliable estimate of the exact GP model. 

Titsias~\cite{titsias2009variational} presents a variational free-energy (VFE) method based on setting the variational parameters as inducing inputs rather than training the inducing inputs as new parameters as subset regression GP approaches.
Using the greedy selection rules, the inducing points are chosen by minimising the Kullback-Leibler divergence between the variational distribution and the exact posterior distribution over the latent variable function values. It results in the evidence lower bound:
\begin{equation}
    \log p(\mathbf{y}) \geq \log \mathcal{N}\left(\mathbf{y} \mid \mathbf{0}, \mathbf{K}_{\mathbf{X} \mathbf{Z}} \mathbf{K}_{\mathbf{Z} \mathbf{Z}}^{-1} \mathbf{K}_{\mathbf{X} \mathbf{Z}}^{\top}+\sigma^{2} \mathbf{I}\right)-\frac{1}{2 \sigma^{2}} \mathrm{tr}\left(\mathbf{K}_{\mathbf{X} \mathbf{X}}-\mathbf{Q}_{\mathbf{X} \mathbf{X}}\right).
\end{equation}
By including extra variational parameters, a more computationally scalable upper bound can be reached~\cite {matthews2016sparse}.
Matthew bridges the gap between the variational inducing-points framework and the more general KL divergence between stochastic processes and applies a standard variational bound $\mathcal{L}=\mathcal{L}_{\text {lower }}+\mathrm{KL}(\mathbf{Q} \| {\mathbf{P}})$. 
where $\mathbf{Q}$ is a variational GP approximation and $\mathbf{P}$ is the true posterior GP.
The sparse variational Gaussian process (SVGP) based on inducing points has been further enhanced recently. In~\cite{shi2020sparse}, a new variational inference framework is proposed. The GP prior is decomposed as the sum of a low-rank approximation using inducing points and a full-rank residual process. This method reduces the complexity of SVGP, which still scales cubically with the number of inducing points. Adam~\cite{adam2020doubly} proposes an improvement inference method for 1-dimensional input space to combine the advantages of both sparse GP approximation and state-space model representation, which results in a novel representation of inducing features as the state space components. The computational complexity grows linearly with the number of data and inducing points.
Moreover, the variational parameters that need to be optimised are also reduced. Hensman and Dutordoir~\cite{hensman2017variational,dutordoir2020sparse} attempt to improve the scalability of SVGP from a different angle. Instead of reducing the cubic complexity with the number of inducing points, it generates the inducing variables by projecting the GP onto the Fourier basis, thus obtaining inducing variables with higher global informativeness on the predictions and reducing the number of inducing points needed for SVGP.
Furthermore, the diagonal covariance matrices are obtained, which fully bypasses the need to compute expensive matrix inverses. For choosing the inducing points, Burt~\cite{burt2019rates} discusses the minimum inducing point number to satisfy the growth of the data size. Uhrenholt~\cite{uhrenholt2021probabilistic} proposes a probabilistic paradigm for balancing the capacity and complexity of sparse Gaussian processes.

\section{Structured Sparse Approximation}
The special structure of the covariance matrix can directly accelerate solving the matrix inversion. It can be achieved through fast matrix-vector multiplication (MVM). When the covariance matrix has some algebraic structures, such as the Kronecker structure, the MVMs can provide massive scalability. Here we will introduce how to use this feature to improve the scalability of the standard GP.
\subsection{Kronecker and Toeplitz Structures}
Assume the multi-dimensional input $\mathbf{x}$ is on the Cartesian grid namely, $\mathbf{x} \in \mathbf{X}_{1} \times \cdots \times \mathbf{X}_{D}$ where $D$ is the dimension of the input space. The product kernel function across grid dimensions can be written as
\begin{equation}
    \mathbf{k}\left(\mathbf{x}_{i}, \mathbf{x}_{j}\right)=\prod_{d=1}^{D} \mathbf{k}\left(\mathbf{x}_{i}^{(d)}, \mathbf{x}_{j}^{(d)}\right).
\end{equation}
Based on properties of the Kronecker product, the inverse of the covariance matrix can be efficiently found from $\mathbf{K}_{1}, \ldots, \mathbf{K}_{d}$:
\begin{equation}
    \mathbf{K}^{-1}=\bigotimes\nolimits_{d=1}^{D} \mathbf{K}_{d}^{-1}.
\end{equation}
These properties can be utilised to reduce the computational resources for the data without noise. The calculation process would be more complex for the data set with noise. Since $\left(\mathbf{K}+\sigma^{2} \mathbf{I}\right)$ cannot be transferred  to Kronecker structure. Specifically, the covariance matrix K in Kronecker structure can be fast eigendecomposed with low computational cost.  
For the one-dimensional training input situation, Toeplitz can be applied, which counteracts Kronecker's shortcoming. When the inputs $\mathbf{x}$ is on a regularly one-dimensional grid and the covariance matrix $K$ is calculated from the stationary covariance kernel, the covariance matrix can be written as $\mathbf{k}\left(\mathbf{x}, \mathbf{x}^{\prime}\right)=\mathbf{k}\left(\mathbf{x}-\mathbf{x}^{\prime}\right)$. The Toeplitz matrices have constant diagonals: $\mathbf{K}_{i, j}=\mathbf{K}_{i+1, j+1}=k\left(\mathbf{x}_{i}-\mathbf{x}_{j}\right)$ and can be embedded into the circulant matrices for the fast MVMs as the Kronecker structure~\cite{wilson2014fast}. It can reduce computational complexity to $\mathcal{O}(n\log n)$ and memory requirements
to $\mathcal{O}(n)$ through the exploitation of Toeplitz structure induced in the covariance matrix, for any stationary kernel.
\subsection{Structured Kernel Interpolation Method}
The inducing point-based approach combined with a sparse approximation is popular since it can be applied outside the data set without requiring any knowledge of its structure. However, in a large data set, when the number of inducing points is $m$ less than $\mathcal{O}\left(\log ^{D} n\right)$, the predictive accuracy cannot be guaranteed~\cite{burt2019rates}. On the other hand, structure exploiting approaches are compelling because they provide incredible gains in scalability, with essentially no losses in predictive accuracy. However, this method requires the input data on the grid, which makes this approach inapplicable.

Wilson~\cite{wilson2015kernel} outputs a SKI method to solve the limitation where the input points are not on the girds. This approach allows us to approximate the $n \times m$ matrix $\mathbf{K}_{\mathbf{X}, \mathbf{Z}}$ by interpolating on the $m \times m$ covariance matrix $\mathbf{K}_{\mathbf{Z}, \mathbf{Z}}$.
For instance, to estimate $k\left(\mathbf{x}_{i}, \mathbf{z}_{j}\right)$ for input point $\mathbf{x}_{i}$ and inducing point $\mathbf{z}_{j}$, Then find two inducing points $\mathbf{z}_{a}$ and $\mathbf{z}_{b}$ which most closely bound $\mathbf{x}_{i}: \mathbf{z}_{a} \leq \mathbf{x}_{i} \leq \mathbf{z}_{b}$ . The kernel function can be written as
\begin{equation}
    \mathbf{k}\left(\mathbf{x}_{i}, \mathbf{z}_{j}\right)=w_{i} \mathbf{k}\left(\mathbf{z}_{a}, \mathbf{z}_{j}\right)+\left(1-w_{i}\right) \mathbf{k}\left(\mathbf{z}_{b}, \mathbf{z}_{j}\right),
\end{equation}
where $w_{i}$ denotes the linear interpolation weights and $\left(1-w_{i}\right)$ represents the relative distances from $\mathbf{x}_{i}$ to points $\mathbf{z}_{a}$ and $\mathbf{z}_{b}$.
Theoretically, all the interpolation strategies can be applied in this method, such as bilinear interpolation or cubic interpolation~\cite{wilson2015kernel}. 

After the interpolation, same as the subset of regression, we set $\mathbf{K}_{\mathbf{X}, \mathbf{Z}} \approx \mathbf{W} \mathbf{K}_{\mathbf{Z}, \mathbf{Z}}$, where $\mathbf{W}$ represent a $n \times m$ interpolation weights matrix. In this case the weight matrix $\mathbf{W}$ can be extremely sparse. For example if it is a local linear interpolation, $\mathbf{W}$ only contains 2 non-zero entries per row.
Substituting this new expression for $\mathbf{K}_{\mathbf{X}, \mathbf{Z}}$ back to the SoR approximation for $\mathbf{K}_{\mathbf{X}, \mathbf{X}}$ the equation can be written as $\mathbf{K}_{\mathbf{X}, \mathbf{X}} \approx \mathbf{W} \mathbf{K}_{\mathbf{Z}, \mathbf{Z}} \mathbf{W}^{\top}$
where $\mathbf{K}_{\mathbf{Z},\mathbf{Z}} \in \mathbb{R}^{m \times m}$ is the kernel matrix for the set of $m$ points on the dense $d$-dimensional grid.
Plugging this approximation into the GP inference. Then, $\mathbf{K}^{-1}\mathbf{y}$ can be solved by linear conjugate gradients, since MVMs only cost $\mathcal{O}\left(n\right)$. The detailed proof and introduction are shown in Chapter 5 of Saatchi~\cite{saatcci2010Gaussian}. Overall, in the Kronecker structure case this equation only requires  $\mathcal{O}\left(Dm^{1+1/D}\right) $ ($D$ is the dimension of the input and $D>1$) computations and $\mathcal{O}\left(n+Dm^{2/D}\right) $ storage~\cite{wilson2014fast}, and the Toeplitz require $\mathcal{O}(m \log m)$ operations and $\mathcal{O}(m)$ storage, where $m$ is the number of grid data points. 
However, based on the conjugate gradient method, the SKI method frame struggles with numerical instabilities in learning kernel hyperparameters and poor test likelihoods. Flaxman~\cite{flaxman2015fast} propose a new scalable Kronecker method for Gaussian processes with non-Gaussian likelihoods. Maddox~\cite{maddox2021iterative} also raises a prescription for the conjugate gradient optimises to correct these issues.

%{\ colour {red}This model also has great applicability, even if the inputs $\mathbf{X}$ do not have any structure, this method allows the naturally created structure in the latent variables $Z$, which can be exploited for greatly accelerated inference and learning. }
%Instead of the gradient descent, linear Conjugate gradient  method can be used here to solve linear system $\left(\mathbf{K}+\sigma^{2} I\right)^{-1} \mathbf{y}$ in likelihood function Eq(4) which 
\section{Hierarchical Matrix-based Approximation}
Besides the inducing point-based methods, another GP method based on the HODLR structure proposed in~\cite{ambikasaran2015fast} has shown a very strong potential to solve the large data. Here we briefly introduce how this structure can be integrated into GP. 

% Minden
\subsection{Structure of the HODLR Matrix}
Firstly we will introduce the HODLR matrices~\cite{aminfar2016fast}. Many mathematicians developed strategies to rearrange the matrix to handle the big, dense covariance matrix. The HODLR matrix, as a sparse representation of the matrix, is one of the appealing structures to solve the covariance matrices. 
The HODLR matrix has many versions or variants with different low-rank approximation methods. However, its main structure remains the same, built via a recursive block partition~\cite{borm2003hierarchical}. This method aims to do the low-rank approximation to the off-diagonal blocks and remains the diagonal parts. According to the k-dimensional tree, sort the data points recursively~\cite{massei2020hm}. Here is an example of a two-level decomposition in the HODLR matrix. A real symmetric matrix $\mathbf{K}\in \mathbb{R}^{n \times n}$ can be decomposed into a two-level HODLR matrix:
\begin{equation}
    \mathbf{K}=\left[\begin{array}{cc}
    \mathbf{K}_{1} & \mathbf{U}_{1} \mathbf{V}_{1}^{T} \\
    \mathbf{V}_{1} \mathbf{U}_{1}^{T} & \mathbf{K}_{2}
    \end{array}\right],
\end{equation}
where the $\mathbf{K}_{1}$ and $\mathbf{K}_{2}$ are the $n/2^{j} \times n/2^{j}$ diagonal block matrices from the original matrix $\mathbf{K}$ and $\mathbf{U}^{(j)}$, $\mathbf{V}^{(j)}$ matrices are $n / 2^{j} \times r$ matrices with $r \ll n$. ${j}$ is the level of decomposition which are 2 in this example and rank $r$ is depends on the desired accuracy of the low-rank approximation. A higher rank results in less precision loss and a higher computational cost. 
The most significant step in constructing a HODLR matrix is compressing the off-diagonal blocks into low-rank small matrices $\mathbf{U}, \mathbf{V}$. Moreover, these ``tall'' and ``thin'' matrices are the most time-consuming step in the factorisation of the HODLR method. The most popular decomposing method is the singular value decomposition (SVD)~\cite{andrews1976singular}. However, SVD is a very expensive computation method. For the square matrix in $n \times n$, the SVD method costs $\mathcal{O}(n^{3})$ since it computes the whole dense matrix. Although it leads to reliable results, applying this method to accelerate the GP is meaningless. Recently, in papers~\cite{aminfar2016fast}, more aggressive strategies have been proposed, such as partially pivoted adaptive cross approximation~\cite{liu2020parallel} or some analytical techniques such as Chebyshev interpolation~\cite{effenberger2012chebyshev}, which can further reduce computational costs to $\mathcal{O}(rn)$. This method allows us to construct the HODLR matrix in a linear scaling with the size of the training data~\cite{ambikasaran2013fast}. Instead of the block-box inducing point model, the low-rank approximation gives a reasonable approximation based on the spectrum of the covariance matrix. The flexible choice of the HODLR matrix level and the low-rank approximation's precision loss allows the model to be more effective in practical problems.

\subsection{Fast Solving Algorithm for Gaussian Process}
After constructing the HODLR matrix, there are different strategies to solve the inverse and determinant by HODLR matrix. The first strategy is the continuous multiplication method~\cite{ambikasaran2015fast,massei2020hm}. 
In the case of a one-level factorization, we can easily write down the computation. Let the matrix $\mathbf{K}$ with the only step in the decomposition is to factor out the terms $\mathbf{K}_{1}, \mathbf{K}_{2}$, giving:
\begin{equation}
    \mathbf{K}=\left[\begin{array}{cc}
\mathbf{K}_{1} & 0 \\
0 &\mathbf{K}_{2}
\end{array}\right]\left[\begin{array}{cc}
\mathbf{I}_{n / 2} & \mathbf{K}_{1}^{-1} \mathbf{U}\mathbf{V}^{T} \\
\mathbf{K}_{2}^{-1} \mathbf{V}\mathbf{U}^{T} & \mathbf{I}_{n / 2}
\end{array}\right].
\end{equation}
The needed computational time is proportional to the inverse of the dense block diagonal matrix to the corresponding rows in the remaining factor. Furthermore, since the matrix $ \mathbf{U}\mathbf{V}^{T}$ is low-rank, so is $\mathbf{K}_{1}^{-1} \mathbf{U}\mathbf{V}^{T}$. Unfortunately, a one-level factorization such as this is still quite expensive: it required the direct inversion of $\mathbf{K}_{1}$ and $\mathbf{K}_{2}$, each of which are $n / 2 \times n / 2$ matrices. The procedure must be recursive across $\log n$ levels to achieve a nearly optimal algorithm.
At first glance, it may look as though the computation of inverse is still expensive and will scale as $\mathcal{O}\left(n^{3} / 8\right)$. However, if the inverses of $\mathbf{K}_{1}, \mathbf{K}_{1}$ are known and $\mathbf{U}, \mathbf{V}$ are low-rank matrices, then the inverse of the full matrix can be computed rapidly using the Sherman-Morrison-Woodbury formula. By apply this formula to diagonal block of each level we will have the factorization equation 
$  \mathbf{K}=\mathbf{K}_{l} \mathbf{K}_{l-1} \cdots \mathbf{K}_{1} \mathbf{K}_{0},$ where the $l$ is the level of the HODLR matrix~\cite{ambikasaran2015fast}. Since each term is block diagonal or a block diagonal low-rank update to the identity matrix, each inverse factorization can be computed in $\mathcal{O}(n \log n)$ time, the total computational to solve the linear system costs $\mathcal{O}\left(n \log ^{2}(n)\right)$.

The second method relies on the Cholesky factorisation of the HODLR matrix~\cite{lyu2022efficient}. Due to low-rank off-diagonal parts, computational costs can be significantly reduced. Ballani~\cite{ballani2016matrices} shows the detailed calculation process for Cholesky decomposition which only results in $\mathcal{O}\left(n \log ^{2}(n)\right)$ costs. Linear systems $\mathbf{A} \mathbf{x}=\mathbf{b}$ with the lower triangular HODLR matrix require additional cost of $\mathcal{O}(n \log (n))$. $\mathcal{O}(n)$ to sum the diagonal elements of the determinant. Besides the fast inverse and determinant, the hyperparameters are learned by maximising the marginal likelihood. Most optimisation methods, such as gradient estimation, require first and second-order derivatives. The Cholesky decomposition with such a tractable structure $\mathbf{A} = \mathbf{L}\mathbf{L}^{\top}$ also saves time for the hyperparameters learning~\cite{litvinenko2019likelihood}. 
\section{Performance Comparison}
To verify all these sparse approximate GP methods, we test them based on the Matlab toolbox GPML and HM-toolbox \cite{rasmussen2010gaussian,massei2020hm}. Figure 1 gives results from the sparse approximate GP methods discussed above. A one-dimensional toy example with $f(x) = 0.02x + \text{sinc}(x)+ \epsilon$  is used in the   comparison where $\epsilon \sim \mathcal{N}(0,0.2)$. The $+$ symbols represent 100 training points.  The top circles represent the initial locations of 10 inducing points, whereas the bottom triangles denote the optimised locations of inducing points. 
\begin{figure}[ht]
    \centering
    \subfigure{
        \includegraphics[width=2.25in]{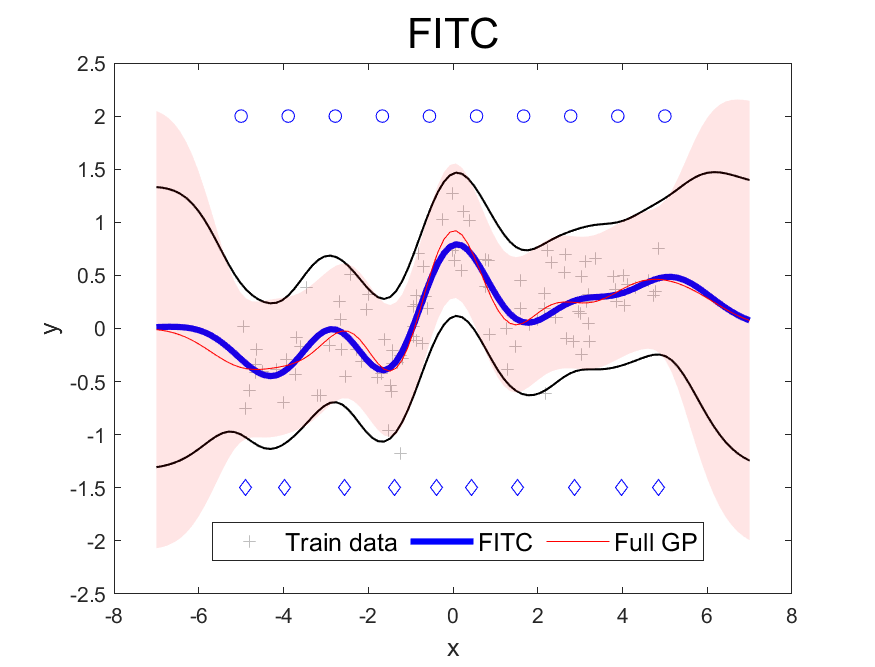}
    }
    \subfigure{
	\includegraphics[width=2.25in]{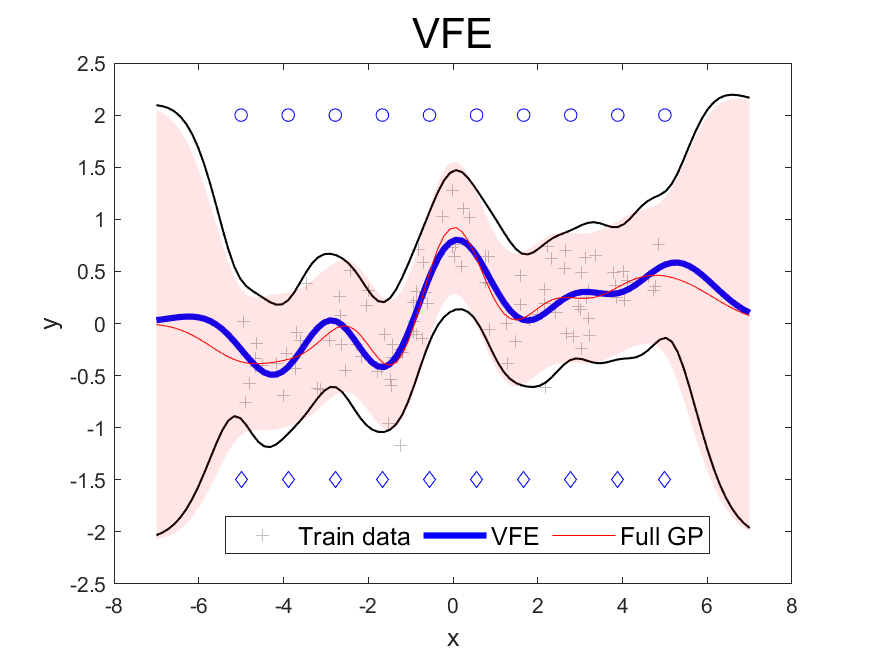}
    }
    \quad 
    \subfigure{
	\includegraphics[width=2.25in]{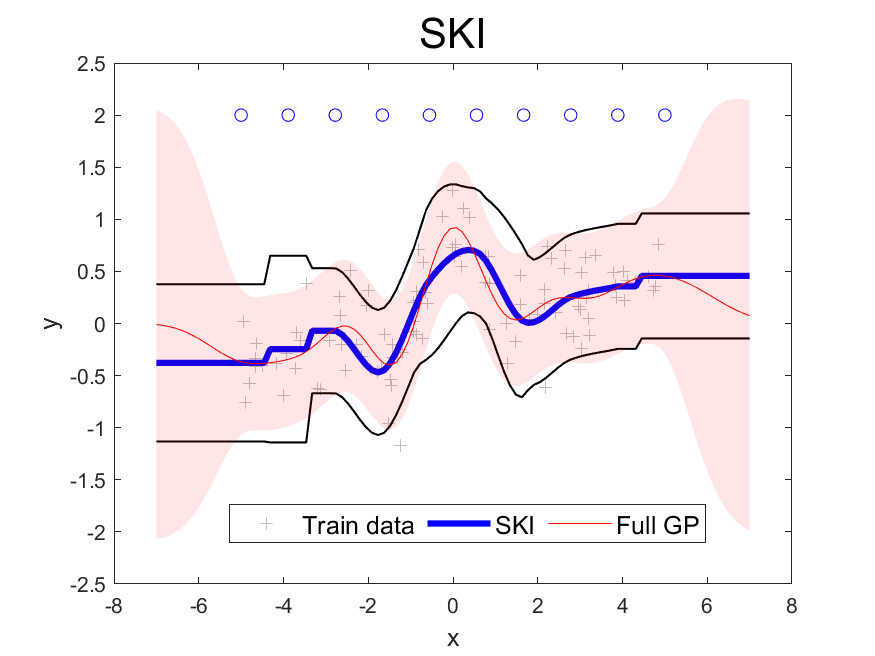}
    }
    \subfigure{
	\includegraphics[width=2.25in]{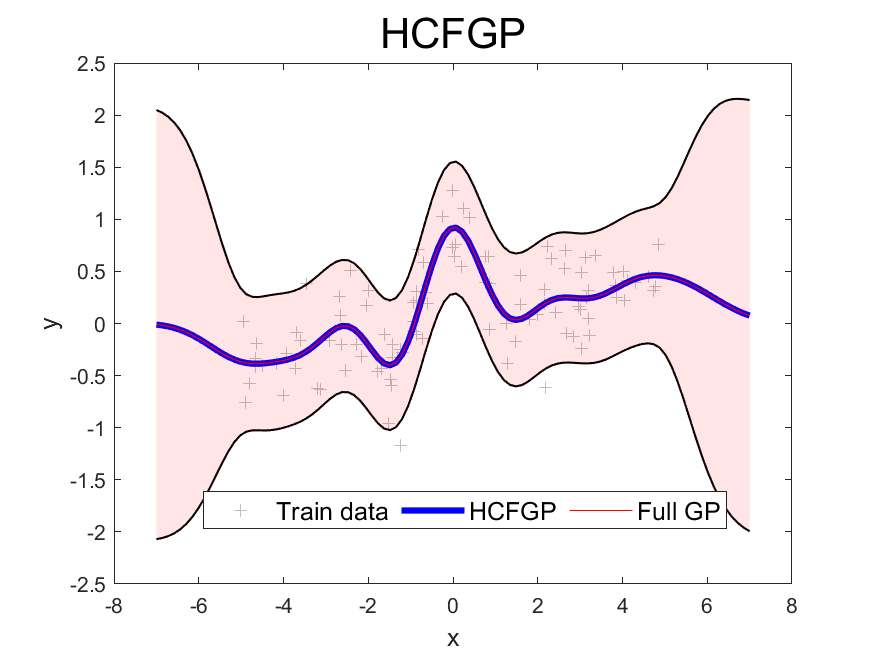}
    }
    \caption{Estimated state and its confidence intervals}
\end{figure}
The dot red curve  represents the predicted mean of the full GP. The red shaded regions visualise the 95\% confidence interval of the predictions of a full GP. The blue curves indicate the predicted mean from the  sparse GP algorithms. The black curves represent the $95\%$ confidence interval for the results from the sparse GP algorithms. 

Figure~1 shows that the FITC and VFE methods give results close to the full GP. However, the FITC method will lead to overfitting due to the greedy selection of the inducing points. The accuracy of these methods relies on the number of the inducing points~\cite{burt2019rates}. Figure 1 also shows that the SKI produces results with some variability due to the interpolation between the insufficient inducing points. The HCFGP has a result that is very close to the full GP in most situations because the only part of the approximation is controllable by the precision loss (rank remained) of the low-rank approximation~\cite{ambikasaran2015fast}. 
\begin{table}[h]
\centering\caption{Computational time of the considered GP methods}
\begin{tabular}{|c|ccccc|}
\hline
\multicolumn{1}{|l|}{} & \multicolumn{5}{c|}{Time(s)}                   \\ \hline
Size of data & \multicolumn{1}{c|}{Full GP} & \multicolumn{1}{c|}{FITC}  & \multicolumn{1}{c|}{VFE}   & \multicolumn{1}{c|}{SKI}   & HCFGP \\ \hline
100                    & \multicolumn{1}{c|}{0.018} & \multicolumn{1}{c|}{0.086} & \multicolumn{1}{c|}{0.019} & \multicolumn{1}{c|}{0.126} & 0.049 \\ \hline
1000                   & \multicolumn{1}{c|}{0.119} & \multicolumn{1}{c|}{0.099} & \multicolumn{1}{c|}{0.030} & \multicolumn{1}{c|}{0.188} & 0.124 \\ \hline
2500                   & \multicolumn{1}{c|}{0.875} & \multicolumn{1}{c|}{0.106} & \multicolumn{1}{c|}{0.042} & \multicolumn{1}{c|}{0.580} & 0.753 \\ \hline
5000                   & \multicolumn{1}{c|}{4.976} & \multicolumn{1}{c|}{0.126} & \multicolumn{1}{c|}{0.051} & \multicolumn{1}{c|}{1.424} & 2.999 \\ \hline
8000         & \multicolumn{1}{c|}{17.805}  & \multicolumn{1}{c|}{0.319} & \multicolumn{1}{c|}{0.120} & \multicolumn{1}{c|}{3.066} & 7.556 \\ \hline
\end{tabular}

\label{tab:my-table}
\end{table}
\noindent The results presented in Table~\ref{tab:my-table} are from 100 repeated Monte Carlo runs. Table~\ref{tab:my-table}  shows the computational time of the compared GP methods. For the big data case, the computation cost of full GP will be cubic growth. FITC and VFE methods takes the least time, which require a small computation cost with a small number of inducing points. The SKI and the HODLR with the Cholesky factorisation Gaussian process (HCFGP) method are slower than other inducing point-based methods due to the time taken for the interpolation and construction of the special matrices. The HCFGP takes less time than SKI when the data size is less than 2500 points. 

\section{Conclusions}
This paper presents recently developed global sparse GP methods, including the classic inducing point-based, structured approximation and new HODLR-based methods. Understanding these methods will help solve the practical high-dimensional big data problems and assess the impact of uncertainties on the developed solutions. The performance of the considered GP methods is evaluated over a different number of data and the efficiency of the factorisation based methods is demonstrated. Factorisation based GP methods can be used in real time applications. 

Another advantage of the considered sparse approximation GP methods is that they can still provide trustworthy solutions and assess the impact of uncertainties on the developed solutions. 

%Constrained

%
%That is exactly the core of the HODLR-based methods, which allows us to understand the approximation methods in an essential view. 
%However these two method offers the more probability to build new a mixture driven approximation method

\vspace{2mm}
\textbf{Acknowledgements:} We are grateful to UK EPSRC for funding this work
through EP/T013265/1 project NSF-EPSRC:ShiRAS.
Towards Safe and Reliable Autonomy in Sensor Driven
Systems and UK Research and Innovation (UKRI)
Trustworthy Autonomous Systems (TAS) programme
[EPSRC Ref: EP/V026747/1]. This work was also supported
by the National Science Foundation under Grant USA NSF
ECCS 1903466. For the purpose of open access, the
authors have applied a Creative Commons Attribution (CC BY)
licence to any Author Accepted Manuscript version arising.

\bibliographystyle{ieeetr}
\bibliography{ref}

\begin{thebibliography}{10}

\bibitem{liu2020Gaussian}
H.~Liu, Y.-S. Ong, X.~Shen, and J.~Cai, ``When {Gaussian} process meets big
  data: A review of scalable {GPs},'' {\em IEEE Transactions on Neural Networks
  and Learning Systems}, vol.~31, no.~11, pp.~4405--4423, 2020.

\bibitem{ki2006Gaussian}
C.~K. Williams and C.~E. Rasmussen, {\em {Gaussian} Processes for Machine
  Learning}.
\newblock MIT Press Cambridge, MA, 2006.

\bibitem{quinonero2005unifying}
J.~Quinonero-Candela and C.~E. Rasmussen, ``A unifying view of sparse
  approximate {Gaussian Process} regression,'' {\em The Journal of Machine
  Learning Research}, vol.~6, pp.~1939--1959, 2005.

\bibitem{quinonero2007approximation}
J.~Quinonero-Candela, C.~E. Rasmussen, and C.~K. Williams, ``Approximation
  methods for {Gaussian} {Processes} regression,'' in {\em Large-scale kernel
  machines}, pp.~203--223, MIT Press, 2007.

\bibitem{shi2011gaussian}
J.~Q. Shi and T.~Choi, {\em Gaussian Process Regression Analysis for Functional
  Data}.
\newblock New York: Chapman and Hall CRC Press, 2011.

\bibitem{ambikasaran2015fast}
S.~Ambikasaran, D.~Foreman-Mackey, L.~Greengard, D.~W. Hogg, and M.~O’Neil,
  ``Fast direct methods for {Gaussian} processes,'' {\em IEEE Transactions on
  Pattern Analysis and Machine Intelligence}, pp.~252--265, 2015.

\bibitem{dai2017efficient}
Z.~Dai, M.~A. {\'A}lvarez, and N.~D. Lawrence, ``Efficient modeling of latent
  information in supervised learning using {Gaussian} {Process}es,'' {\em
  CoRR}, 2017.

\bibitem{snelson2006sparse}
E.~Snelson and Z.~Ghahramani, ``Sparse {Gaussian} {Process}es using
  pseudo-inputs,'' {\em Advances in Neural Information Processing Systems},
  vol.~18, p.~1257, 2006.

\bibitem{saatcci2010Gaussian}
Y.~Saat{\c{c}}i, R.~D. Turner, and C.~E. Rasmussen, ``{Gaussian} {Process}
  change point models,'' in {\em Proc. of the International Conference on
  Machine Learning}, 2010.

\bibitem{blei2017variational}
D.~M. Blei, A.~Kucukelbir, and J.~D. McAuliffe, ``Variational inference: A
  review for statisticians,'' {\em Journal of the American Statistical
  Association}, vol.~112, no.~518, pp.~859--877, 2017.

\bibitem{titsias2009variational}
M.~Titsias, ``Variational learning of inducing variables in sparse {Gaussian}
  {Processes}es,'' in {\em Artificial Intelligence and Statistics}, PMLR, 2009.

\bibitem{matthews2016sparse}
A.~G. d.~G. Matthews, J.~Hensman, R.~Turner, and Z.~Ghahramani, ``On sparse
  variational methods and the {Kullback-Leibler} divergence between stochastic
  {Processes}es,'' in {\em Artificial Intelligence and Statistics},
  pp.~231--239, PMLR, 2016.

\bibitem{shi2020sparse}
J.~Shi, M.~Titsias, and A.~Mnih, ``Sparse orthogonal variational inference for
  {Gaussian} processes,'' in {\em Proc. of the International Conf. on AI and
  Statistics}, pp.~1932--1942, PMLR, 2020.

\bibitem{adam2020doubly}
V.~Adam, S.~Eleftheriadis, A.~Artemev, N.~Durrande, and J.~Hensman, ``Doubly
  sparse variational {Gaussian} processes,'' in {\em International Conference
  on Artificial Intelligence and Statistics}, pp.~2874--2884, PMLR, 2020.

\bibitem{hensman2017variational}
J.~Hensman, N.~Durrande, A.~Solin, {\em et~al.}, ``Variational fourier features
  for {Gaussian} processes.,'' {\em J. Mach. Learn. Res.}, vol.~18, no.~1,
  pp.~5537--5588, 2017.

\bibitem{dutordoir2020sparse}
V.~Dutordoir, N.~Durrande, and J.~Hensman, ``Sparse gaussian processes with
  spherical harmonic features,'' in {\em Proc. of the International Conference
  on Machine Learning}, pp.~2793--2802, PMLR, 2020.

\bibitem{burt2019rates}
D.~Burt, C.~E. Rasmussen, and M.~Van Der~Wilk, ``Rates of convergence for
  sparse variational {Gaussian} process regression,'' in {\em Proc. of the
  International Conference on Machine Learning}, pp.~862--871, PMLR, 2019.

\bibitem{uhrenholt2021probabilistic}
A.~K. Uhrenholt, V.~Charvet, and B.~S. Jensen, ``Probabilistic selection of
  inducing points in sparse gaussian processes,'' in {\em Uncertainty in
  Artificial Intelligence}, pp.~1035--1044, PMLR, 2021.

\bibitem{wilson2014fast}
A.~G. Wilson, E.~Gilboa, J.~P. Cunningham, and A.~Nehorai, ``Fast kernel
  learning for multidimensional pattern extrapolation.,'' in {\em NIPS},
  pp.~3626--3634, 2014.

\bibitem{wilson2015kernel}
A.~Wilson and H.~Nickisch, ``Kernel interpolation for scalable structured
  {Gaussian} {Processes}es ({KISS-GP}),'' in {\em Proc. of the International
  Conf. on Machine Learning}, pp.~1775--1784, PMLR, 2015.

\bibitem{flaxman2015fast}
S.~Flaxman, A.~Wilson, D.~Neill, H.~Nickisch, and A.~Smola, ``Fast kronecker
  inference in gaussian processes with non-gaussian likelihoods,'' in {\em
  Proc. of the International Conference on Machine Learning}, pp.~607--616,
  PMLR, 2015.

\bibitem{maddox2021iterative}
W.~J. Maddox, S.~Kapoor, and A.~G. Wilson, ``When are iterative gaussian
  processes reliably accurate?,'' {\em arXiv preprint arXiv:2112.15246}, 2021.

\bibitem{aminfar2016fast}
A.~Aminfar, S.~Ambikasaran, and E.~Darve, ``A fast block low-rank dense solver
  with applications to finite-element matrices,'' {\em Journal of Computational
  Physics}, vol.~304, pp.~170--188, 2016.

\bibitem{borm2003hierarchical}
S.~B{\"o}rm, L.~Grasedyck, and W.~Hackbusch, ``Hierarchical matrices,'' {\em
  Lecture notes}, vol.~21, p.~2003, 2003.

\bibitem{massei2020hm}
S.~Massei, L.~Robol, and D.~Kressner, ``hm-toolbox: {Matlab} software for
  {HODLR} and {HSS} matrices,'' {\em SIAM J. on Scient. Comp.}, vol.~42, no.~2,
  pp.~C43--C68, 2020.

\bibitem{andrews1976singular}
H.~Andrews and C.~Patterson, ``Singular value decomposition (svd) image
  coding,'' vol.~24, no.~4, pp.~425--432, 1976.

\bibitem{liu2020parallel}
Y.~Liu, W.~Sid-Lakhdar, E.~Rebrova, P.~Ghysels, and X.~S. Li, ``A parallel
  hierarchical blocked adaptive cross approximation algorithm,'' {\em The
  International J. of High Performance Computing Applications}, vol.~34, no.~4,
  pp.~394--408, 2020.

\bibitem{effenberger2012chebyshev}
C.~Effenberger and D.~Kressner, ``Chebyshev interpolation for nonlinear
  eigenvalue problems,'' {\em BIT Numerical Mathematics}, vol.~52, no.~4, 2012.

\bibitem{ambikasaran2013fast}
S.~Ambikasaran, {\em Fast algorithms for dense numerical linear algebra and
  applications}.
\newblock Stanford University, 2013.

\bibitem{lyu2022efficient}
C.~Lyu, X.~Liu, and L.~Mihaylova, ``Efficient factorisation-based {Gaussian}
  process approaches for online tracking,'' in {\em Proc. of the 25th
  International Conf. on Information Fusion}, IEEE, 2022.

\bibitem{ballani2016matrices}
J.~Ballani and D.~Kressner, ``Matrices with hierarchical low-rank structures,''
  in {\em Exploiting hidden structure in matrix computations: algorithms and
  applications}, pp.~161--209, Springer, 2016.

\bibitem{litvinenko2019likelihood}
A.~Litvinenko, Y.~Sun, M.~G. Genton, and D.~E. Keyes, ``Likelihood
  approximation with hierarchical matrices for large spatial datasets,'' {\em
  Computational Statistics \& Data Analysis}, vol.~137, pp.~115--132, 2019.

\bibitem{rasmussen2010gaussian}
C.~E. Rasmussen and H.~Nickisch, ``Gaussian processes for machine learning
  (gpml) toolbox,'' {\em The Journal of Machine Learning Research}, vol.~11,
  pp.~3011--3015, 2010.

\end{thebibliography}

\end{document}